\documentclass[11pt,twoside,english]{article}
\usepackage{amsmath}
\usepackage{amssymb}
\usepackage{graphicx}
\usepackage{esint}

\providecommand{\tabularnewline}{\\}

\usepackage{cite}

\begin{document}
\global\long\def\pst{\hspace{1em}*{1.5em}}

\global\long\def\rigmark{\em Journal of Russian Laser Research}
 \global\long\def\lemark{\em Volume 30, Number 5, 2009}

\global\long\def\be{ 
\begin{equation}
\end{equation}
 }
 \global\long\def\ee{{equation}}
 \global\long\def\bm{\boldmath}
 \global\long\def\ds{{\displaystyle }}
 \global\long\def\bea{ 
\begin{eqnarray}
\end{eqnarray}
 }
 \global\long\def\eea{{eqnarray}}
 \global\long\def\ba{\begin{array}{c}
 \end{array}}
 \global\long\def\ea{{array}}
 \global\long\def\arcsinh{\mathop{\rm arcsinh}\nolimits}
 \global\long\def\arctanh{\mathop{\rm arctanh}\nolimits}
 \global\long\def\bc{\begin{center}\end{center}}
 \global\long\def\ec{{center}}

\thispagestyle{plain}

\label{sh}


\begin{center}
\textbf{\Large }%
\begin{tabular}{c}
\textbf{\Large SHANNON ENTROPY AS A MEASURE OF UNCERTAINTY }\tabularnewline
\textbf{\Large IN POSITIONS AND MOMENTA}\tabularnewline
\end{tabular}\textbf{ }
\par\end{center}

\bigskip{}

\bigskip{}

\begin{center}
\textbf{\L{}ukasz Rudnicki$^{*}$ }
\par\end{center}

\medskip{}

\begin{center}
\textit{Center for Theoretical Physics, Polish Academy of Sciences}\\
\textit{ Aleja Lotnik{\'o}w 32/46, PL-02-668 Warsaw, Poland}
\par\end{center}

\begin{center}
\smallskip{}

\par\end{center}

\begin{center}
$^{*}$Corresponding author e-mail:~~~rudnicki@cft.edu.pl\\
 
\par\end{center}
\begin{abstract}
\noindent This paper is prepared as a contribution to the proceedings
after the 12th ICSSUR/Feynfest Conference held in Foz do Igua{\c c}u (Brazil)
from 2 to 6 May 2011. In the first part I briefly report the topic
of entropic uncertainty relations for position and momentum variables.
Then I investigate the discrete Shannon entropies related to the case
of finite number of detectors set to measure probability distributions
in position and momentum spaces. I derive an uncertainty relation
for the sum of the Shannon entropies which generalizes previous approaches
[\textit{Phys. Lett.} \textbf{103 A}, 253 (1984)] based on an
infinite number of detectors (bins).
\end{abstract}
\medskip{}


\section{Introduction}

Entropic uncertainty relations for position and momentum, or other
canonically conjugated variables have been derived a long time ago.
Initial investigations \cite{BBM} were devoted to continuous Shannon
entropies. Further generalizations took into account, in a spirit
of Deutsch \cite{Deutsch} and Maassen-Uffink \cite{mu} results,
the accuracy of measuring devices \cite{partovi,IBB0,IBBr,Pramana,sen}
and impurity of a quantum state \cite{VVD}. Recent 12th ICSSUR/Feynfest
Conference showed that the topic of entropic uncertainty relations
including experimental accuracies is important in the task of entanglement
detection in quantum optics \cite{Antonio,Saboia}.

This paper is organized as follows. In a further part of the present
section I point out some aspects related to entropic uncertainty relations,
following the rephrased Heisenberg sentence \cite{sen} \textit{,,the
more information we have about the position, the less information
we can acquire about the momentum and vice versa''}
and an observation made by Peres \cite{peres} \textit{,,The
uncertainty relation such as $\sigma_{x}\sigma_{p}\geq\hbar/2$ is
not a statement about the accuracy of our measuring instruments''}.
It the second section I generalize an approach presented in \cite{IBB0}
to the case of finite number of measuring devices (detectors). 

\subsection{Entropic uncertainty relations and continuous Shannon entropy}

I would like to start with the famous Bialynicki-Birula-Mycielski
entropic uncertainty relation of the form \cite{BBM}
\begin{equation}
-\int_{-\infty}^{\infty}dx\rho\left(x\right)\ln\rho\left(x\right)-\int_{-\infty}^{\infty}dp\tilde{\rho}\left(p\right)\ln\tilde{\rho}\left(p\right)\geq1+\ln\pi\hbar,\label{BBM}
\end{equation}
where $\rho\left(x\right)=\left|\psi\left(x\right)\right|^{2}$ and
$\tilde{\rho}\left(p\right)=\left|\tilde{\psi}\left(p\right)\right|^{2}$
are probability distributions in position and momentum spaces respectively.
Wave functions in both spaces are related to each other by the Fourier
transform
\begin{equation}
\tilde{\psi}\left(p\right)=\frac{1}{\sqrt{2\pi\hbar}}\int_{-\infty}^{\infty}dxe^{-ipx/\hbar}\psi\left(x\right).
\end{equation}
The introduction of \cite{BBM} starts as follows:

\medskip{}

\textit{,,The purpose of this paper is to derive a new stronger
version of the Heisenberg uncertainty relation in wave mechanics.
This new uncertainty relation has a simple interpretation in terms
of information theory. It is also closely related to newly discovered
logarithmic Sobolev inequalities.''}

\medskip{}
Information theory enters to the inequality (\ref{BBM}) by the notion
of continuous Shannon entropies $S^{(x)}=-\int_{-\infty}^{\infty}dx\rho\left(x\right)\ln\rho\left(x\right)$
and $S^{(p)}=-\int_{-\infty}^{\infty}dp\tilde{\rho}\left(p\right)\ln\tilde{\rho}\left(p\right)$.
Connection with the logarithmic Sobolev inequality can be recognized
with the help of the reversed logarithmic Sobolev inequality \cite{Sobol}
which, for the probability distribution function (PDF) $f\left(z\right)$
defined on $\Omega\subset\mathbb{R}$, reads 
\begin{equation}
\int_{\Omega}dzf\left(z\right)\ln f\left(z\right)\geq-\frac{1}{2}\ln\left[2\pi e\sigma_{z}^{2}\left(\Omega\right)\right].\label{LogSobol}
\end{equation}
The variance is defined as usual, $\sigma_{z}^{2}\left(\Omega\right)=\int_{\Omega}dz\, z^{2}f\left(z\right)-\left(\int_{\Omega}dz\, zf\left(z\right)\right)^{2}$.
We shall use the inequality (\ref{LogSobol}) independently for the
position and momentum variables and obtain the stronger version of
the Heisenberg uncertainty relation, announced in \cite{BBM}
\begin{equation}
\sigma_{x}\sigma_{p}\geq\frac{\hbar}{2}\exp\left(S^{(x)}+S^{(p)}-1-\ln\pi\hbar\right)\geq\frac{\hbar}{2}.
\end{equation}

\subsubsection{Continuous Shannon entropy as a measure of information}

It seems to be widely accepted that the continuous Shannon entropy
is also a good measure of information since it is a relative of Shannon
information entropy. However, this statement is not completely true.
To prove that let me recall the definition of the Shannon entropy
of a set of probabilities $\left\{ P_{i}\right\} $
\begin{equation}
H^{\left(P\right)}=-\sum_{i}P_{i}\ln P_{i}.\label{Shannon}
\end{equation}
We have two important properties of the Shannon entropy (\ref{Shannon}):
\begin{enumerate}
\item Since the probabilities $P_{i}$ are dimensionless the Shannon entropy
$H^{\left(P\right)}$ is also dimensionless.
\item From the property $0\leq P_{i}\leq1$ we know that $H^{\left(P\right)}\geq0$.
\end{enumerate}
These two properties are essential for the information-like interpretation
of the Shannon entropy because information can be neither negative
nor expressed in any physical units. Unfortunately the continuous
Shannon entropy does not possess these two properties. For instance
the unit of the entropy $S^{\left(x\right)}$ (in SI units) is the
logarithm of meter. This makes impossible to check if the continuous
Shannon entropy is positive or negative. On the other hand if we introduce
some length (momentum) scale in order to measure $S^{\left(x\right)}$
($S^{\left(p\right)}$) then we still cannot fulfill the second property
since $\rho\left(x\right)$ ($\tilde{\rho}\left(p\right)$) can be
greater than $1$ and make the continuous entropy negative.

\subsection{Discrete Shannon entropies including experimental accuracies}

An idea that helps to overcome the difficulties which I have mentioned
above (related to the continuous entropies) introduces a partition
of the real line into bins of equal width (see Fig. 1). 

\begin{figure}
\centering{}\includegraphics[bb=0bp 80bp 214bp 214bp]{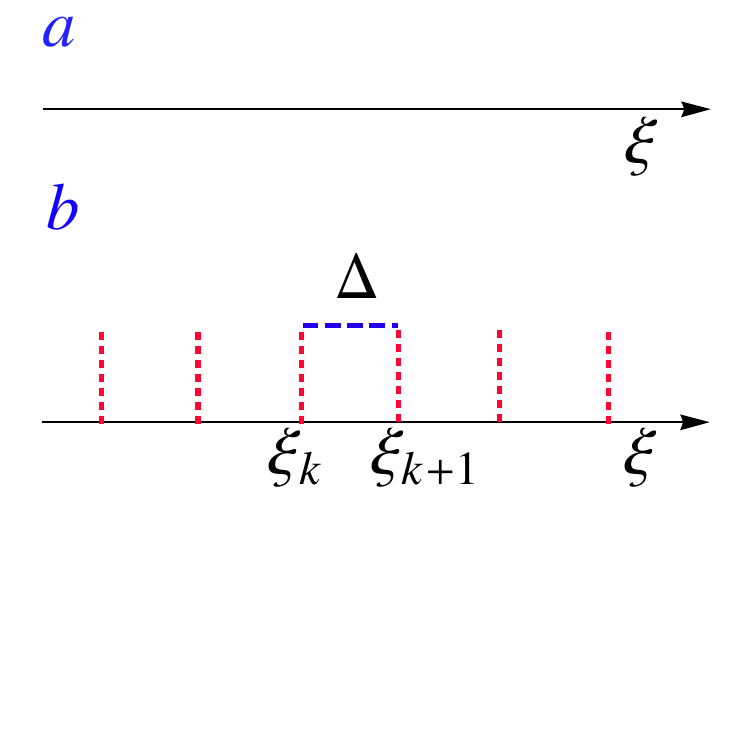}\label{fig1}\caption{Partition of the real line into bins.}
\end{figure}

As a result the continuous probability distribution $f\left(\xi\right)$
is replaced by a discrete distribution $P_{k}$, ($k\in\mathbb{Z}$)
\begin{equation}
P_{k}=\int_{\xi_{k}}^{\xi_{k+1}}d\xi f\left(\xi\right).
\end{equation}
The bins are of equal width, thus $\forall_{k}\xi_{k+1}-\xi_{k}=\Delta$.
This idea was for the first time introduced by Partovi \cite{partovi}
who noticed that $\Delta$ shall be interpreted as a \textit{{},,resolution
of the measuring device''}. 

\subsubsection{A remark about translational invariance}

The requirement that all bins have to have the same width $\Delta$
is a strong constraint on $\xi_{k}$. In fact $\xi_{k}$ must be of
the form
\begin{equation}
\xi_{k}=\xi_{0}+k\Delta,\label{con}
\end{equation}
so there is only one free parameter $\xi_{0}$ to be chosen. One might
expect that different values of the parameter $\xi_{0}$ are equivalent
to different possible choices of the central point ($\xi=0$) of the
used coordinate. In the first Partovi's paper \cite{partovi} and
later publications \cite{IBB0,IBBr,WW} the easiest possible choice
$\xi_{0}=0$ was made. In this choice the probability distributions
and the Shannon entropies are ($\delta x$ and $\delta p$ are experimental
accuracies for positions and momenta respectively):
\begin{equation}
\mathcal{Q}_{k}=\int_{k\delta x}^{\left(k+1\right)\delta x}dx\rho\left(x\right),\qquad\mathcal{P}_{l}=\int_{l\delta p}^{\left(l+1\right)\delta p}dp\tilde{\rho}\left(p\right),\label{en1}
\end{equation}
\begin{equation}
H^{(x)}=-\sum_{k=-\infty}^{\infty}\mathcal{Q}_{k}\ln\mathcal{Q}_{k},\qquad H^{(p)}=-\sum_{l=-\infty}^{\infty}\mathcal{P}_{l}\ln\mathcal{P}_{l}.\label{en2}
\end{equation}

In this moment I would like to point out that the entropies defined
in (\ref{en1}, \ref{en2}) are not correct measures of information
(because of the choice $\xi_{0}=0$). In order to realize that one
shall investigate the limit of large coarse graining (large experimental
accuracies) $\delta x\rightarrow\infty$ or $\delta p\rightarrow\infty$.
In these limits performed measurements tell us nothing about the quantum
state - our information is $0$, thus, we shall expect $\lim_{\delta x\rightarrow\infty}H^{(x)}=$$\lim_{\delta p\rightarrow\infty}H^{(p)}=0$.
But in fact, we have:
\begin{equation}
\lim_{\delta x\rightarrow\infty}\mathcal{Q}_{k}=\begin{cases}
\int_{0}^{\infty}dx\rho\left(x\right) & \textrm{for }k=0\\
\int_{-\infty}^{0}dx\rho\left(x\right) & \textrm{for }k=-1
\end{cases},\qquad\lim_{\delta p\rightarrow\infty}\mathcal{P}_{l}=\begin{cases}
\int_{0}^{\infty}dp\tilde{\rho}\left(p\right) & \textrm{for }l=0\\
\int_{-\infty}^{0}dp\tilde{\rho}\left(p\right) & \textrm{for }l=-1
\end{cases}.
\end{equation}
As a result $\lim_{\delta x\rightarrow\infty}H^{(x)}$ ($\lim_{\delta p\rightarrow\infty}H^{(p)}$)
have state-dependent values that vary between $0$ if the state is
localized in $\mathbb{R}_{+}$ or $\mathbb{R}_{-}$ (in positions
or momenta) and $\ln2$ if the state is symmetric.

In \cite{comment,sen} we captured this ambiguity and performed the
redefinition of the probability distributions (\ref{en1}) in the
following way:
\begin{equation}
\mathcal{Q}_{k}\longmapsto q_{k}=\int_{\left(k-1/2\right)\delta x}^{\left(k+1/2\right)\delta x}dx\rho\left(x\right),\qquad\mathcal{P}_{l}\longmapsto p_{l}=\int_{\left(l-1/2\right)\delta p}^{\left(l+1/2\right)\delta p}dp\tilde{\rho}\left(p\right).\label{en1-1}
\end{equation}
It is equivalent to the choice $\xi_{0}=-\Delta/2$ in the construction
(\ref{con}) and means that the center of the coordinate $\xi=0$
lays in the middle of the central bin. In the previous choice the
center of the coordinate was the border point between two bins.

\subsubsection{Entropic uncertainty relations and recent results}

Similarly to (\ref{BBM}) uncertainty relations for $H^{(x)}+H^{(p)}$
were found a long time \cite{partovi,IBB0}. The stronger one \cite{IBB0}
reads
\begin{equation}
H^{(x)}+H^{(p)}\geq-\ln\left(\frac{\delta x\delta p}{e\pi\hbar}\right)=\mathcal{B}.\label{Ibb}
\end{equation}
Since the Shannon entropies (\ref{en2}) have been correctly defined
(are positive and dimensionless) it is obvious that $H^{(x)}+H^{(p)}\geq0$.
Thus, for $\delta x\delta p\geq e\pi\hbar$ the relation (\ref{Ibb})
becomes trivially satisfied and in fact does not give an optimal lower
bound (the bound is optimal, and saturated by a Gaussian distribution,
only in the limit $\delta x\rightarrow0$ and $\delta p\rightarrow0$).
Some attempts \cite{WW} to find a better lower bound failed completely
(for details see \cite{comment} - the comment on \cite{WW}). Recently
\cite{LR,PRA} we derived the bound

\begin{figure}
\centering{}\includegraphics{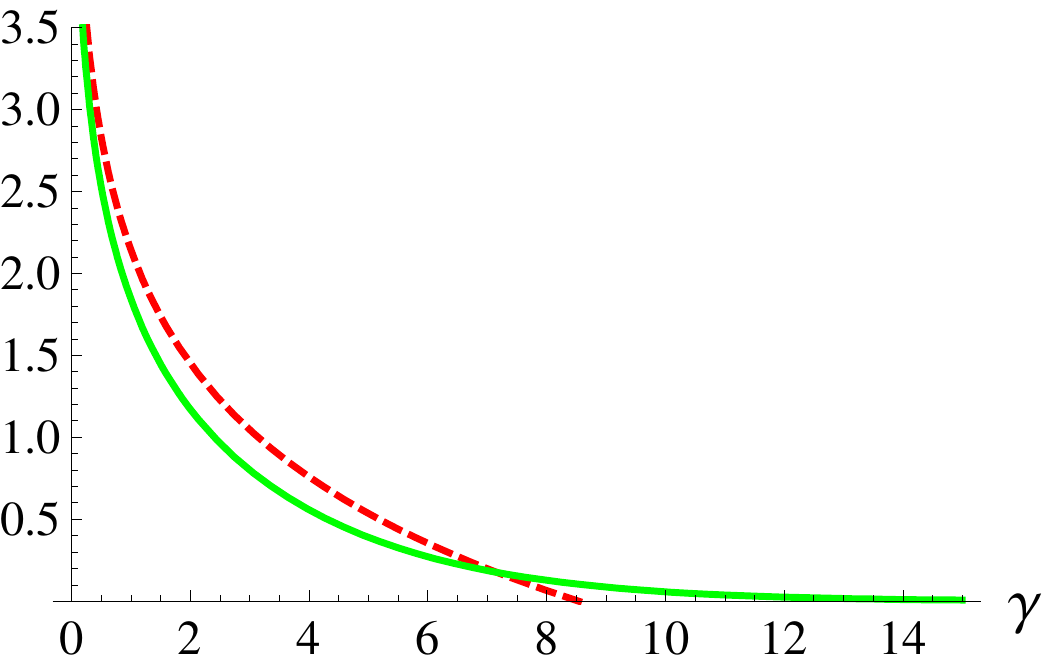}\label{fig2}\caption{Comparison between two bounds in (\ref{bound new}). Red/dashed curve
represents $\mathcal{B}\left(\gamma\right)$ while green curve represents
$\mathcal{R}\left(\gamma\right)$; $\gamma=\delta x\delta p/\hbar$.}
\end{figure}
\begin{equation}
H^{(x)}+H^{(p)}\geq\max\left(\mathcal{B},\mathcal{R}\right),\qquad\mathcal{R}=-2\ln\left[\sqrt{\frac{\delta x\delta p}{2\pi\hbar}}R_{00}\left(\frac{\delta x\delta p}{4\hbar},1\right)\!\right],\label{bound new}
\end{equation}
which is always positive, but still not optimal (cf. Fig. 2).
Function $R_{00}\left(\xi,\eta\right)$ %
\footnote{in the Wolfram Mathematica's notation it reads $\mathtt{SpheroidalS1[0,0,\xi,\eta]}$.%
} is one of the radial prolate spheroidal wave functions of the first
kind \cite{abr}.

\section{Entropy based on a finite number of detectors}

In the definition (\ref{en2}) the summation over $k$ ($l$) goes
from $-\infty$ to $+\infty$. That property in a language of an experiment
means that there are infinitely many measuring devices (detectors)
covering an infinite number of bins. Of course in realistic experiments
only a finite number of bins can be covered, thus, we shall describe
the measurement's result with the help of the finite Shannon entropies,
defined as follows: 

\begin{equation}
H_{M}^{(x)}=-\sum_{k=-M}^{M}q_{k}\ln q_{k},\qquad H_{N}^{(p)}=-\sum_{l=-N}^{N}p_{l}\ln p_{l}.
\end{equation}
These definitions include $2M+1$ detectors set to measure the probability
distribution in positions and $2N+1$ detectors for momenta. We have
the limits $\lim_{M\rightarrow\infty}H_{M}^{(x)}=H^{(x)}$ and $\lim_{N\rightarrow\infty}H_{N}^{(p)}=H^{(p)}$.
Since all terms in the sums in (\ref{en2}) are non-negative we know
that $H_{M}^{(x)}\leq H^{(x)}$ and $H_{N}^{(p)}\leq H^{(p)}$. Thus,
we cannot expect that the uncertainty relations (\ref{Ibb}) or (\ref{bound new})
will be satisfied for $H_{M}^{(x)}+H_{N}^{(p)}$. Moreover we can
always find many states localized far away from the detectors (with
$\left|\left\langle x\right\rangle \right|\gg\left(M+1/2\right)\delta x$
and $\left|\left\langle p\right\rangle \right|\gg\left(N+1/2\right)\delta p$),
for example
\begin{equation}
\psi\left(x\right)=\left(\frac{1}{\pi\sigma^{2}}\right)^{1/4}e^{ip_{0}\left(x-x_{0}/2\right)/\hbar}\exp\left[-\frac{\left(x-x_{0}\right)^{2}}{2\sigma^{2}}\right],\;\Longrightarrow\quad\rho\left(x\right)=\frac{1}{\sqrt{\pi}\sigma}\exp\left[-\frac{\left(x-x_{0}\right)^{2}}{\sigma^{2}}\right],
\end{equation}
with $x_{0}=\left\langle x\right\rangle $ and $p_{0}=\left\langle p\right\rangle $.
This inconvenience is directly related to the translational invariance
which was broken when we restricted the number of used bins. Of course
an experimentalist knows where the measured state is and can properly
choose the coordinate to assure that $\left\langle x\right\rangle =0=\left\langle p\right\rangle $. 

Looking at Fig. 2 we can realize that if $\delta x\delta p/\hbar\lesssim7.167$
then the previous lower bound (\ref{Ibb}) dominates. I would like
to work in this regime and, although the finite number of bins makes
us unable to find a state independent lower bound for $H_{M}^{(x)}+H_{N}^{(p)}$,
I would like to find a state-dependent correction to the formula (\ref{Ibb}).
I shall start in the same way as was done in \cite{IBB0} and use
the integral Jensen inequality (I will do the calculations only for
the position variable since in the momentum case everything shall
go the same way)
\begin{align}
H_{M}^{(x)}\geq- & \int_{-\left(M+1/2\right)\delta x}^{\left(M+1/2\right)\delta x}dx\rho\left(x\right)\ln\left[\rho\left(x\right)\delta x\right]\\
= & B_{x}+q_{\infty}\ln\left(\delta x\right)+\left(\int_{-\infty}^{-\left(M+1/2\right)\delta x}+\int_{\left(M+1/2\right)\delta x}^{\infty}\right)dx\rho\left(x\right)\ln\rho\left(x\right),\nonumber 
\end{align}
where I have introduced the following notation:
\begin{equation}
B_{x}=-\ln\left(\delta x\right)+S^{(x)}\qquad q_{\infty}=\left(\int_{-\infty}^{-\left(M+1/2\right)\delta x}+\int_{\left(M+1/2\right)\delta x}^{\infty}\right)dx\rho\left(x\right).
\end{equation}
The Shannon entropy, because of the logarithmic function, collects
information from all moments of the probability distribution. A convenient
method to reduce the state-dependent input can be based on the reversed
logarithmic Sobolev inequality (\ref{LogSobol}). We shall choose
$f\left(z\right)=\rho\left(z\right)/q_{\infty}$ and $\Omega_{M}=\left]-\infty,-\left(M+1/2\right)\delta x\right]\cup\left[\left(M+1/2\right)\delta x,\infty\right[$
and obtain

\begin{equation}
\left(\int_{-\infty}^{-\left(M+1/2\right)\delta x}+\int_{\left(M+1/2\right)\delta x}^{\infty}\right)dx\rho\left(x\right)\ln\rho\left(x\right)\geq-\frac{q_{\infty}}{2}\ln\left(\frac{2\pi e}{q_{\infty}^{2}}\sigma_{x}^{2}\left(\Omega_{M}\right)\right).\label{19}
\end{equation}
Since we expect that the state fulfills $\left\langle x\right\rangle \approx0\approx\left\langle p\right\rangle $
we can, without any significant loss, simplify the right hand side
of (\ref{19})
\begin{equation}
\sigma_{x}^{2}\left(\Omega_{M}\right)\leq\frac{\left\langle x^{2}\right\rangle _{M}}{q_{\infty}},\qquad\left\langle x^{2}\right\rangle _{M}=\left(\int_{-\infty}^{-\left(M+1/2\right)\delta x}+\int_{\left(M+1/2\right)\delta x}^{\infty}\right)dx\, x^{2}\rho\left(x\right).
\end{equation}
After this step we obtain the inequality
\begin{equation}
H_{M}^{(x)}\geq B_{x}+R\left(q_{\infty},\left\langle x^{2}\right\rangle _{M}\right),\qquad R\left(\eta,\Lambda\right)=\frac{\eta}{2}\ln\left(\frac{\left(\delta x\right)^{2}\eta^{3}}{2\pi e\Lambda}\right),\qquad\eta\in\left[0,1\right],\quad\Lambda\geq0,
\end{equation}
where the $R$ function depends only on the ,,$0$th moment''
(the norm $q_{\infty}$) and the ,,$2$nd moment'' ($\left\langle x^{2}\right\rangle _{M}$)
of the function $\rho\left(x\right)$ restricted to $\Omega_{M}$.
Since the moments of the function are independent we can find $\min_{q_{\infty}}R\left(q_{\infty},\left\langle x^{2}\right\rangle _{M}\right)$
keeping $\left\langle x^{2}\right\rangle _{M}$ constant. This step
will reduce the state dependent input only to the one quantity $\left\langle x^{2}\right\rangle _{M}$.
In order to find the minimum we shall calculate the following derivatives
\begin{equation}
\frac{\partial}{\partial\eta}R\left(\eta,\Lambda\right)=\frac{3}{2}+\frac{1}{2}\ln\left(\frac{\left(\delta x\right)^{2}\eta^{3}}{2\pi e\Lambda}\right),\qquad\frac{\partial^{2}}{\partial\eta^{2}}R\left(\eta,\Lambda\right)=\frac{3}{2\eta}.
\end{equation}
Since for $\eta\in\left[0,1\right]$ the second derivative is positive
we have only one global minimum $\eta_{min}=\left(\sqrt{2\pi\Lambda}/e\delta x\right)^{2/3}$.
On the other hand when $2\pi\Lambda\geq\left(e\delta x\right)^{2}$
the minimum lays outside the domain of $\eta$. Thus, since $R$ is
a decreasing function on the interval $\left[0,\eta_{min}\right]$
we shall modify $\eta_{min}$ and write 
\begin{equation}
\eta_{min}\left(\Lambda\right)=\min\left\{ \left(\frac{\sqrt{2\pi\Lambda}}{e\delta x}\right)^{2/3},1\right\} .
\end{equation}
Using this result we obtain that: 
\begin{equation}
\begin{cases}
R\left(q_{\infty},\left\langle x^{2}\right\rangle _{M}\right)\geq-3\left(\frac{\sqrt{\pi\left\langle x^{2}\right\rangle _{M}}}{2e\delta x}\right)^{2/3} & \textrm{ when }\left\langle x^{2}\right\rangle _{M}<\frac{\left(e\delta x\right)^{2}}{2\pi}\\
R\left(q_{\infty},\left\langle x^{2}\right\rangle _{M}\right)\geq\ln\left(\frac{\delta x}{\sqrt{2\pi e\left\langle x^{2}\right\rangle _{M}}}\right) & \textrm{ when }\left\langle x^{2}\right\rangle _{M}\geq\frac{\left(e\delta x\right)^{2}}{2\pi}
\end{cases}.
\end{equation}
If we assume that the numbers $M$ ($N$) are sufficiently large,
so that $2\pi\left\langle x^{2}\right\rangle _{M}<\left(e\delta x\right)^{2}$
and $2\pi\left\langle p^{2}\right\rangle _{N}<\left(e\delta p\right)^{2}$
we find the following uncertainty relation
\begin{equation}
H_{M}^{(x)}+H_{N}^{(p)}\geq-\ln\left(\frac{\delta x\delta p}{e\pi\hbar}\right)-3\left(\frac{\sqrt{\pi\left\langle x^{2}\right\rangle _{M}}}{2e\delta x}\right)^{2/3}-3\left(\frac{\sqrt{\pi\left\langle p^{2}\right\rangle _{N}}}{2e\delta p}\right)^{2/3}>-\ln\left(\frac{\delta x\delta p}{e\pi\hbar}\right)-3.
\end{equation}
This bound gives a nontrivial limitation when $\delta x\delta p/\hbar\leq\pi e^{-2}\approx0.425$,
what it this case is equivalent to $\left\langle x^{2}\right\rangle _{M}\left\langle p^{2}\right\rangle _{N}<\hbar^{2}/4$.
In a general case the uncertainty relation for the sum of the entropies
is $H_{M}^{(x)}+H_{N}^{(p)}\geq\mathcal{L}$, where:
\begin{equation}
\mathcal{L}=\begin{cases}
-\ln\left(\frac{\delta x\delta p}{e\pi\hbar}\right)-3\left(\frac{\sqrt{\pi\left\langle x^{2}\right\rangle _{M}}}{2e\delta x}\right)^{2/3}-3\left(\frac{\sqrt{\pi\left\langle p^{2}\right\rangle _{N}}}{2e\delta p}\right)^{2/3} & \textrm{when }\left\langle x^{2}\right\rangle _{M}<\frac{\left(e\delta x\right)^{2}}{2\pi},\;\textrm{and }\left\langle p^{2}\right\rangle _{N}<\frac{\left(e\delta p\right)^{2}}{2\pi}\\
-3\left(\frac{\sqrt{\pi\left\langle x^{2}\right\rangle _{M}}}{2e\delta x}\right)^{2/3}-\ln\left(\frac{\delta x\sqrt{2\left\langle p^{2}\right\rangle _{N}}}{\sqrt{e\pi}\hbar}\right) & \textrm{when }\left\langle x^{2}\right\rangle _{M}<\frac{\left(e\delta x\right)^{2}}{2\pi},\;\textrm{and }\left\langle p^{2}\right\rangle _{N}\geq\frac{\left(e\delta p\right)^{2}}{2\pi}\\
-3\left(\frac{\sqrt{\pi\left\langle p^{2}\right\rangle _{N}}}{2e\delta p}\right)^{2/3}-\ln\left(\frac{\delta p\sqrt{2\left\langle x^{2}\right\rangle _{M}}}{\sqrt{e\pi}\hbar}\right) & \textrm{when }\left\langle x^{2}\right\rangle _{M}\geq\frac{\left(e\delta x\right)^{2}}{2\pi},\;\textrm{and }\left\langle p^{2}\right\rangle _{N}<\frac{\left(e\delta p\right)^{2}}{2\pi}\\
-\ln\left(\frac{2\sqrt{\left\langle x^{2}\right\rangle _{M}\left\langle p^{2}\right\rangle _{N}}}{\hbar}\right) & \textrm{when }\left\langle x^{2}\right\rangle _{M}\geq\frac{\left(e\delta x\right)^{2}}{2\pi},\;\textrm{and }\left\langle p^{2}\right\rangle _{N}\geq\frac{\left(e\delta p\right)^{2}}{2\pi}
\end{cases}.\label{final}
\end{equation}

\section{Summary}

I have briefly presented a current status in the topic of entropic
uncertainty relations for position and momentum variables. I have
pointed out ambiguities appearing in the scientific literature. Finally
I have investigated the case of finite number of detectors in an approach
using discrete Shannon entropies including experimental accuracies
and derived nontrivial, state-dependent lower bound generalizing previous
results.

\section*{Acknowledgments}

I am specially indebted to Iwo Bialynicki-Birula who inspired and
supported all my efforts in the topic of entropic uncertainty relations.
I would like to thank Stephen P. Walborn, Fabricio Toscano, Luiz Davidovich
and all other members of Faculty of Physics in Federal University
of Rio de Janeiro (Instituto de F{\'{i}}sica in Universidade Federal do
Rio de Janeiro) for their hospitality after the 12th ICSSUR/Feynfest
Conference. This research was partly supported by a grant from the
Polish Ministry of Science and Higher Education for the years 2010\textendash{}2012.

\end{document}